\documentclass[nojss]{jss}
\usepackage{tabularx}
\usepackage{amsmath,amssymb,amscd}
\usepackage{mathtools}
\usepackage{subfigure}
\usepackage{natbib}
\def \C{\mathcal{C}}

\def\sT{\mbox{\tiny$T$}}

\def \I{\mathcal{I}}

\DeclareMathOperator*{\argmax}{\arg\!\max}

\newtheorem{thm}{Theorem}

\newtheorem{prop}{Proposition}

\author{Dang Trung Kien, Neo Han Wei and Sanjay Chaudhuri\\National University of Singapore, DBS Bank and National University of Singapore
        }
\title{\pkg{elhmc}: An R Package for Hamiltonian Monte Carlo Sampling in Bayesian Empirical Likelihood}

\Plainauthor{Dang Trung Kien, Neo Han Wei, Sanjay Chaudhuri} 
\Plaintitle{elhmc: An R Package for Hamiltonian Monte Carlo Sampling in Bayesian Empirical Likelihood} 
\Shorttitle{elhmc} 

\Abstract{
In this article, we describe a {\tt R} package for sampling from an empirical likelihood-based posterior using a Hamiltonian Monte Carlo method. Empirical likelihood-based methodologies have been used in Bayesian modeling of many problems of interest in recent times.  
This semiparametric procedure can easily combine the flexibility of a non-parametric distribution estimator together with the interpretability of a parametric model.  The model is specified by estimating equations-based constraints.  
Drawing an inference from a Bayesian empirical likelihood (BayesEL) posterior is challenging.  The likelihood is computed numerically, so no closed expression of the posterior exists.  Moreover, for any sample of finite size, the support of the likelihood is non-convex, which hinders the fast mixing of many Markov Chain Monte Carlo (MCMC) procedures.
It has been recently shown that using the properties of the gradient of log empirical likelihood, one can devise an efficient Hamiltonian Monte Carlo (HMC) algorithm to sample from a BayesEL posterior.  
 The package requires the user to specify only the estimating equations, the prior, and their respective gradients.  An MCMC sample drawn from the BayesEL posterior of the parameters, with various details required by the user is obtained.
}

\Plainkeywords{Empirical likelihood, Bayesian empirical likelihood, Logposterior gradient, Hamiltonian Monte Carlo}

\Address{
  Dang Trung Kien\\ 
  Department of Biostatistics\\
  Saw Swee Hock School of Public Health\\
  National University of Singapore\\
  E-mail: \email{stadtk@nus.edu.sg},\\
  Neo Han Wei\\
  DBS Bank, Singapore\\
  E-mail:\email{nhanwei@me.com},\\
  and\\
  Sanjay Chaudhuri\\
  Department of Statistics and Applied Probability\\
  Faculty of Science\\
  National University of Singapore\\
  Blk S16, Level 7, 6 Science Drive 2, Singapore\\
  E-mail: \email{stasc@nus.edu.sg}\\
}

\begin{document}

\section{Introduction}
Empirical likelihood has several advantages over the traditional parametric likelihood. Even though, a correctly specified parametric likelihood is usually the most efficient for parameter estimation, semi-parametric methods like empirical likelihood which uses a non-parametric estimate of the underlying distribution are often more efficient when the model is specified wrong.  
Empirical likelihood incorporates parametric model based information as constraints in estimating the underlying distribution, which makes the parametric estimates interpretable.  Furthermore, it allows easy incorporation of known additional information not involving the parameters in the analysis.   

Bayesian empirical likelihood (BayesEL) methods employ empirical likelihood in Bayesian paradigm. Given some information about the model parameters in form of a prior distribution and estimating equations obtained from the model, a likelihood is constructed from a constrained empirical estimate of the underlying distribution. The prior is then used to define a posterior based on this estimated likelihood.  
Inference on the parameter is drawn based on samples generated from the posterior distribution.    

BayesEL methods are quite flexible and has been found useful in many areas of statistics.  The examples include, small area estimation, quantile regression, analysis of complex survey data etc.  

BayesEL procedures however require an efficient Markov Chain Monte Carlo (MCMC) procedure to sample from the resulting posterior.  It turns out that such an procedure is not easily specified.  For many parameter values, it may not be feasible to compute the constrained empirical distribution function and the likelihood is estimated to be zero.  That is, the estimated likelihood is not supported over the whole space. Moreover, this support is non-convex and impossible to determine in most cases. 
Thus a naive random walk MCMC would quite often propose parameters outside the support and get stuck. 


Many authors have encountered this problem in frequentist applications. Such "empty set" problems are quite common \citep{grendar2009empty} and become more frequent in problems with large number of parameters \citep{bergsma2012empty}. 
Several authors \citep{chen2008adjusted,emerson2009calibration,liu2010adjusted} have suggested addition of extra observations generated from the available data designed specifically to avoid empty sets. They show that such observations can be proposed without changing the asymptotic distribution of the corresponding Wilk's statistics. 
Some authors (\citep{tsao2013extending,tsao2013empirical,tsaoFu2014}) have used a transformation so that the contours of resultant empirical likelihood could be extended beyond the feasible region. However, In most Bayesian applications the data is limited, for which the asymptotic arguments have little use. 

With the availability of user friendly software packages like {\tt STAN} \citep{stan2017}, gradient assisted MCMC methods like Hamiltonian Monte Carlo (HMC) are becoming increasingly popular in Bayesian computation.  When the estimating equations are smooth with respect to the parameters, gradient based methods would have a huge advantage in sampling from a BayesEL posterior. 
This is because \citet{chaudhuriMondalTeng2017} have shown that under mild conditions, the gradient of the log-posterior would diverge to infinity at the boundary of its support. 
Due to this phenomenon, if a HMC chain approaches the boundary of the posterior support it would be reflected towards its centre.

There is no software to implement HMC sampling from a BayesEL posteriors with smooth estimating equation and priors.  We describe such a library called {\tt elhmc} written for {\tt R} platform. The main function in the library only requires the user to specify the estimating equations, prior and respectively their hessian and gradient with respect to the parameters as functions.  Outputs with user specified degree of details can be obtained.

The rest of the article is structured as follows. We start with the theoretical background behind the software package. In section \ref{sec:theory} we first define the empirical likelihood and construct a Bayesian empirical likelihood from it.  
Next part of this section is devoted to a review of the properties of the log empirical likelihood gradient.  A review of HMC method with special emphasis on BayesEL sampling is provided next (Section \ref{sec:hmc}).  
Section \ref{sec:package} mainly contains the description of the {\tt elhmc} library.  Some illustrative examples with artificial and real data sets are presented in Section \ref{sec:examples}.

\section{Theoretical Background}\label{sec:theory}
\subsection{Basics of Bayesian Empirical Likelihood}
Suppose $x=(x_1,\ldots,x_n)\in \mathbb{R}^p$ are $n$ observations from a distribution $F^0$ depending on a parameter vector $\theta=(\theta^{(1)}, \ldots,\theta^{(d)})\in\Theta\subseteq \mathbb{R}^d$. We assume that both $F^0$ and the true parameter value $\theta^0$ are unknown. However, certain smooth functions $g(\theta,x)=\left(g_1(\theta,x),\ldots,g_q(\theta,x)\right)^T$ are known to satisfy
\begin{equation}\label{smoothfun}
E_{F^0}[g(\theta^0,x)]=0.
\end{equation}

Additionally, information about the parameter are available in the form of a prior density $\pi(\theta)$ supported on $\Theta$. We assume that it is neither possible nor desirable to specify $F^0$ in a parametric form.  On the other hand, it is not beneficial to estimate $F^0$ completely non-parametrically without taking into account the information from \eqref{smoothfun} in the estimation procedure. 

Empirical likelihood provides a semi-parametric procedure to estimate $F^0$, by incorporating information contained in \eqref{smoothfun}. A likelihood can be computed from the estimate.  Moreover, if some information about the parameter is available in the form of a prior distribution, the same likelihood can be employed to derive a posterior of the parameter given the observations.

Let $F\in\mathcal{F}_{\theta}$ be a distribution function depending on the parameter $\theta$. Empirical likelihood is the maximum of the ``non-parametric likelihood" 
\begin{equation}\label{eqn2}
L(F)=\prod_{i=1}^n \{F(x_i)-F(x_i-)\}
\end{equation}
over $\mathcal{F}_\theta$, $\theta\in\Theta$, under constraints depending on $g(\theta,x)$.

More specifically, by defining $\omega_i=F(x_i)-F(x_i-)$, the empirical likelihood for $\theta$ is defined by,
\begin{equation}\label{eqn3}
L(\theta)\coloneqq\max_{\omega\in\mathcal{W}_{\theta}}\prod_{i=1}^n \omega_i
\end{equation}
where
\[
\mathcal{W}_{\theta}=\Big\{\omega: \sum_{i=1}^n\omega_i g(\theta,x_i)=0\Big\}\cap\Delta_{n-1}
\]
and $\Delta_{n-1}$ is the $n-1$ dimensional simplex, i.e. $\omega_i\geq 0$, $\forall i$ and $\sum_{i=1}^n\omega_i=1$.  For any $\theta$, if the problem in \eqref{eqn3} is infeasible, i.e. $\mathcal{W}_{\theta}=\emptyset$, we define $L(\theta)\coloneqq 0$.

Using the empirical likelihood $L(\theta)$ and the prior $\pi(\theta)$ we can define a posterior as:
\begin{equation}\label{eqn4}
\Pi(\theta|x)=\frac{L(\theta)\pi(\theta)}{\int L(\theta)\pi(\theta) d\theta}\propto L(\theta)\pi(\theta). 
\end{equation}

In Bayesian empirical likelihood (BayesEL), $\Pi(\theta|x)$ is used as the posterior to draw inferences on the parameter.

Returning back to \eqref{eqn3} above, suppose we denote:
\begin{equation}\label{eqn5}
\hat{\omega}\left(\theta\right)=\argmax_{\omega\in\mathcal{W}_{\theta}}\prod_{i=1}^n \omega_i.\hspace{1in}\text{$\Bigl( \text{ i.e. } L(\theta)=\prod^n_{i=1}\hat{\omega}_i\left(\theta\right)\Bigr)$}
\end{equation}
Each $\hat\omega_i\geq 0$ if and only if the origin in $\mathbb{R}^q$ can be expressed as a convex combination of $g(\theta,x_1),\ldots,g(\theta,x_n)$. Otherwise, the optimisation problem is infeasible and $\mathcal{W}_{\theta}=\emptyset$.  Furthermore, when $\hat{\omega}_i>0$, $\forall i$ is feasible, the solution $\hat{\omega}$ of \eqref{eqn5} is unique.  

The estimate of $F^0$ is given by:
\[
\hat{F}^0(x)=\sum_{i=1}^n\hat{\omega}_i(\theta)1_{\{x_i\leq x\}}.
\]
The distribution $\hat{F}^0$ is a step function with a jump of $\hat{\omega}_i(\theta)$ on $x_i$.  If $\mathcal{W}_{\theta}=\Delta_{n-1}$, i.e. no information about $g(\theta,x)$ is present, it easily follows that $\hat{\omega}_i(\theta)=n^{-1}$, for each $i=1$, $2$, $\ldots$, $n$ and $\hat{F}^0$ is the well-known empirical distribution function. 



By construction, $\Pi(\theta|x)$ can only be computed numerically. No analytic form is available.  Inferences are drawn through the observations from $\Pi(\theta|x)$ sampled using Markov chain Monte Carlo techniques.


Adaptation of Markov chain Monte Carlo methods to BayesEL applications pose several challenges. First of all, it is not possible to determine the full conditional densities in a closed form.  So techniques like Gibb's sampling \citep{geman1984stochastic} cannot be used. In most cases, random walk Metropolis procedures, with carefully chosen step sizes are attempted. However, the nature of the support of $\Pi(\theta|x)$, which we discuss in details below, makes the choice of an appropriate step size extremely difficult.

Provided that the prior is positive over whole $\Theta$, which is true in most applications, the support of $\Pi(\theta|x)$ is a subset of the support of the likelihood $L(\theta)$ which can be defined as:
\begin{equation}\label{support}
\Theta_1=\left\{\theta: L(\theta)>0\right\}.
\end{equation}
Thus, the efficiency of MCMC algorithm would depend on $\Theta_1$ and the behaviour of $\Pi(\theta|x)$ on it. 

By definition $\Theta_1$ is closely connected to the set
\begin{equation}\label{convexhull}
\C(\theta,x)=\left\{\sum_{i=1}^n\omega_ig(\theta,x_i) \Big|\omega\in \Delta_{n-1}\right\},
\end{equation}
which is the closed convex hull of the $q$ dimensional vectors $G(x,\theta)=\{g(\theta,x_i),\ldots,g(\theta,x_n)\}$ in $\mathbb{R}^q$.  Suppose $\C^0(\theta,x)$ and $\partial \C(\theta,x)$ are respectively the interior and boundary of $\C(\theta,x)$.  By construction, $\C(\theta,x)$ is a convex polytope.  Since the data $x$ is fixed, 
the set $\C(\theta,x)$ is a set-valued function of $\theta$.  For any $\theta\in\Theta$, the problem in \eqref{eqn3} is feasible (i.e. $\mathcal{W}_{\theta}\ne\emptyset$) if and only if the origin of $\mathbb{R}^q$, denoted by $0_q$, is in $\C(\theta,x)$.  That is, $\theta\in\Theta_1$ if and only if, the same $0_q\in\C^0(\theta,x)$.  
It is not possible to determine $\Theta_1$ in general.  The only way is to check if for any potential $\theta$, the origin $0_q$ is in $\C^0(\theta,x)$.  There is no quick numerical way to check the latter either. Generally, an attempt is made to solve \eqref{eqn3}.  Existence of such a solution indicates that $\theta\in L(\theta)$.

Examples show \citep{chaudhuriMondalTeng2017} that even for simple problems $\Theta_1$ may not be a convex set.  Designing an efficient random walk Markov chain Monte Carlo algorithm on a potentially non-convex support is an extremely challenging task.  
Unless the step-sizes and the proposal distributions are adapted well to the proximity of the current position to the boundary of $\Theta_1$, the chain may repeatedly propose values outside the likelihood support and as a result converge very slowly.  
Adaptive algorithms like the one proposed by \citet{haario1999adaptive} do not tackle the non-convexity problem well.  

Hamiltonian Monte Carlo methods solve well known equations of motions from classical mechanics to propose new values of $\theta\in\Theta$.  Numerical solutions of these equations of motion are dependent on the gradient of the log posterior. 
The norm of the gradient of the log empirical likelihood used in BayesEL procedures diverges near the boundary of $\Theta_1$.  This property makes the Hamiltonian Monte Carlo procedures very efficient for sampling a BayesEL posterior.  It ensures that once in $\Theta_1$, the chain would rarely step outside the support and repeatedly sample from the posterior.


\subsection{A Review of Some Properties of the Gradient of Log Empirical Likelihood}\label{sec:elprop}

Various properties of log-empirical likelihood have been discussed in literature. However, the properties of its gradients with respect to the model parameters are relatively unknown. Our main goal in this section is to review the behaviour of gradients of log-empirical likelihood on the support of the empirical likelihood.   We only state the relevant results here. The proofs of these results can be found in \citet{chaudhuriMondalTeng2017}.

Recall that, the support $\Theta_1$ can only be specified by checking if $0_q\in\C^0(x,\theta_0)$ for each individual $\theta_0\in\Theta$.  If for some $\theta_0\in\Theta$,  the origin lie on the boundary of $\C(x,\theta_0)$, i.e. $0_q\in\partial \C(x,\theta_0)$, 
the problem in \eqref{eqn3} is still feasible, however, $L\left(\theta_0\right)=0$ and the solution of \eqref{eqn5} is not unique. Below we discuss how, under mild conditions, for any $\theta_0\in\Theta$, for a large subset $S\subseteq\partial \C(x,\theta_0)$, if $0_q\in S$, 
the absolute value of least one component of the gradient of $\log\left(L\left(\theta_0\right)\right)$ would be large.  


 Before we proceed, we make the following assumptions:

\begin{enumerate}
\item[(A0)] $\Theta$ is an open set.  \label{A0}

\item[(A1)] $g$ is a continuously differentiable function of  $\theta$ in $\Theta$, $q \le d$ and $\Theta_1$ is non-empty. \label{A1}

\item[(A2)] The sample size $n > q$.  The matrix $G(x, \theta)$  has full row rank for any $\theta \in \Theta$.  

\item[(A3)]  For any fixed $x$, let  $\nabla g(x_i,\theta)$  be the  $q \times d$ Jacobian matrix for any $\theta \in \Theta$.  Suppose $w=(w_1,\ldots, w_n)\in\Delta_{n-1}$ and there are at least $q$ elements of $w$ that are greater than $0$. Then, for any $\theta \in \Theta$,  the matrix  $\sum_{i=1}^n w_i \nabla g(x_i,\theta)$ has full row rank.
\end{enumerate}

Under the above assumptions, several results about the log empirical likelihood and its gradient can be deduced \citep{chaudhuriMondalTeng2017}.

First of all, since the properties of the gradient of the log empirical likelihood at the boundary of the support is of interest, some topological properties of the support needs to investigated. Under the standard topology of $\mathbb{R}^q$, since $\C(x,\theta)$ is a convex polytope with finite number of faces and extreme points, using the smoothness of $g$, it is easy to see that, 
for any $\theta_0\in\Theta_1$ one can find an real number $\delta$, such that the open ball centred at $\theta_0$ with radius $\delta$ is contained in $\Theta_1$. That is, (see \citet[Proposition $1$]{chaudhuriMondalTeng2017} for a more formal proof) the following result holds.  

\begin{prop}\label{prop1} Under assumptions A0, A1 and A2,  $\Theta_1$ is an open subset of $\Theta$. 
\end{prop}
Since $\Theta_1$ is an open set.  The boundary $\partial\Theta_1$ of $\Theta_1$ is not contained in $\Theta_1$.  Now, in view of Proposition \ref{prop1} and the definition of $\Theta_1$ and $\C(x,\theta)$ the following result immediately follows:

\begin{prop}\label{prop2}
Let $\theta^{(0)}\in\partial\Theta_1$ and lies within $\Theta$. Then the primal problem \eqref{eqn3} is feasible at $\theta^{(0)}$ and $0_q \in \partial \C(x,\theta^{(0)})$. 
\end{prop}

Our main objective is to study the utility of Hamiltonian Monte Carlo methods for drawing samples from a BayesEL posterior. The sampling scheme will produce a sequence of sample points in $\theta^{(k)}\in\Theta_1$.  It would be efficient as long as $\log\left(\theta^{(k)}\right)$ is large.  The sampling scheme could potentially become inefficient if some $\theta^{(k)}$ 
is close to the boundary $\partial\Theta_1$.  Thus, it is sufficient to consider the properties of the log empirical likelihood and its gradient along such a sequence converging to a point  $\theta^{(0)}\in\partial\Theta_1$.    

From Proposition \ref{prop2} it is evident that when $\theta^{(0)} \in \partial \Theta_1$ the problem in \eqref{eqn3} is feasible but the likelihood $L\left(\theta^{(0)}\right)$ will always be zero and \eqref{eqn5} will not have a unique solution.  Since $\C(x,\theta^{(0)})$ is a polytope, and $0_q$ lies on one of its faces,  
 there exists a subset $\I_0$ of the observations and $0$ belongs to the interior of the convex hull generated by all $g(x_i,\theta^{(0)})$ for $i \in \I_0$, it follows from the supporting hyperplane theorem \citep{boyd2004convex} that there exists an unit vector $a\in \mathbb{R}^q$ such that
\[
a^{\sT} g(x_i, \theta^{(0)}) =0 \quad \mbox{for} \quad i \in \I_0, \qquad\text{and}\qquad a^{\sT} g(x_{i}, \theta^{(0)}) >0 \quad \mbox{for} \quad i \in \I_0^c.
\]
From some algebraic manipulation it easily follows that any $\omega\in\mathcal{W}_{\theta^{(0)}}$ as defined in \eqref{eqn3} must satisfy, 
\[\omega_i=0 \quad \mbox{for} \quad i \in \I_0^c \qquad\text{and}\qquad \omega_i>0 \quad \mbox{for} \quad i \in \I_0. 
\]  

The solution of \eqref{eqn5} i.e. $\hat{w}(\theta)$ is smooth for all $\theta\in\Theta_1$ \citep{qin1994empirical}.  As $\theta^{(k)}$ converge to $\theta^{(0)}$, the properties of $\hat{w}(\theta^{(k)})$ need to be considered.  To that goal, we first make a specific choice of $\hat{w}(\theta^{(0)})$.   

First we consider a restriction of problem \eqref{eqn5} to $\I_0$.  

\begin{equation}\label{submax}
\hat\nu(\theta)  =\argmax_{\nu\in\mathcal{V}_\theta} \prod_{i\in\I_0} \nu_i
\end{equation}
where
\[
\mathcal{V}_\theta=\left\{\nu: \sum_{i\in \I_0}\nu_i g(x_i,\theta)=0\right\}\cap\Delta_{|\I_0|-1}.
\]
We now define
\[
\hat  \omega_i(\theta^{(0)}) = \hat\nu(\theta^{(0)}), \quad i \in \I_0 \quad \mbox{and} \quad \hat  \omega_i(\theta^{(0)}) = 0, \quad i \in \I_0^c,
\]
and
\[
L(\theta^{(0)})= \prod_{i=1}^n \hat  \omega_i(\theta^{(0)}).  
\]   

Since $\theta^{(0)}$ is in the interior of $\I_0$, the problem \eqref{submax} has a unique solution.  For each $\theta^{(k)}\in\Theta_1$, $\hat{\omega}(\theta^{(k)})$ is continuous taking values in a compact set. 
 Thus as $\theta^{(k)}$ converges to $\theta^{(0)}$, $\hat{\omega}(\theta^{(k)})$ converges to a limit. Furthermore, this limit is a solution of \eqref{eqn5} at $\theta^{(0)}$. However, counterexamples show \citep{chaudhuriMondalTeng2017} that the limit may not be $\hat  \omega_i(\theta^{(0)})$ as defined above.  
That is, the vectors $\hat{\omega}(\theta^{(k)})$ do not extend continuously to the boundary $\partial\Theta_1$ as a whole.  The following result about the components $\hat{\omega}_i(\theta^{(k)})$, for $i\in \I_0^c$ and the likelihood can be proved.

 \begin{thm}\label{continuity1}
Let  $\{ \theta^{(k)}\}$, $k=1,2, \ldots$, be a sequence of points in $\Theta_1$ such that $\theta^{(k)}$ converges to a boundary point $\theta^{(0)}$ of $\Theta_1$. Assume that $\theta^{(0)}$ lies within $\Theta$.  Let  $\I_0$ be the subset of $\{1,2,\ldots,n\}$ such that $\hat\omega_i(\theta^{(0)}) >0$ for all $i \in \I_0$. It then follows that 
\begin{enumerate}
\item $\lim_{k\to\infty}\hat\omega_i(\theta^{(k)}) = \hat \omega_i(\theta^{(0)})   = 0$, for all $i \in \I_0^c$.
\item $\lim_{k\to\infty}L(\theta^{(k)})=L(\theta^{(0)})=0$.
\end{enumerate}
\end{thm} 

Theorem \ref{continuity1} shows that the components of $\hat\omega(\theta^{(k)})$ which are zero in $\hat\omega(\theta^{(0)})$ are continuously extendable.  Furthermore the likelihood is continuous at $\theta^{(0)}$. Such, however is not true for the components $\hat{\omega}_i\left(\theta^{(k)}\right)$, $i\in\I_0$ for which $\hat{\omega}_i\left(\theta^{(k)}\right)\ne 0$.

Since the set $\C(x,\theta)$ is a convex polytope in $\mathbb{R}^q$ the maximum dimension of any of its face is $q-1$, which would have exactly $q$ extreme points. Furthermore, any face with smaller dimension can be expressed as an intersection of such $q-1$ dimensional faces.  Next we define
\begin{align}\label{Theta_2}
  &\C(x_{\I},\theta) = \left\{\sum_{i \in \I} \omega_i g(x_i,\theta)\, \Big|\, \omega\in \Delta_{|\I|-1}\right\}\text{ and }\partial\Theta_1^{(q-1)}= \Bigl\{  \theta:  0 \in  \C^0(x_{\I},\theta)  \mbox{ for some }& \I \Bigr.\nonumber\\
  &\bigl.\mbox{ such that $\C(x_{\I},\theta)$ has exactly $q$ extreme points }\Bigr\}  \cap\partial\Theta_1.
\end{align}

 Thus $ \partial\Theta_1^{(q-1)}$ is the set of all boundary points $\theta^{(0)}$ of $\Theta_1$ such that $0$ belongs to a $(q-1)$-dimensional face of the convex hull $\mathcal{C}(x,\theta^{(0)})$. Now for any $\theta^{(0)}\in \partial\Theta_1^{(q-1)}$, there is a unique set of weight $\nu\in\Delta_{|\I|-1}$ such that, $\sum_{i\in\I}\nu_ig\left(x_i,\theta^{(0)}\right)=0$.  
That is, the set of feasible solutions of \eqref{submax} is a singleton set.  This, after taking note that $\hat{\omega}$ takes values in a compact set, an argument using convergent subsequences, implies that for any sequence $\theta^{(k)}\in\Theta_1$ converging to $\theta^{(0)}$, the whole vector $\hat{\omega}\left(\theta^{(k)}\right)$ converges to $\hat{\omega}\left(\theta^{(0)}\right)$.  
That is the whole vector $\hat{\omega}\left(\theta^{(k)}\right)$ extends continuously to $\hat{\omega}\left(\theta^{(0)}\right)$.





We now consider the behaviour of the gradient of the log empirical likelihood near the boundary of $\Theta_1$.  First, note that, for any $\theta \in \Theta_1$, the gradient of the log empirical likelihood  is  given by
\[
\nabla \log L(\theta)  =  -n\sum_{i=1}^n \hat \omega_i(\theta)    \hat{\lambda}(\theta)^{\sT}  \nabla g(x_i,\theta).
\]
where $\hat{\lambda}(\theta)$ is the estimated Lagrange multiplier satisfying the equation 
\begin{equation}\label{eq:lagmult}
\sum_{i=1}^n \frac{g(x_i,\theta)}{\left\{1+ \hat\lambda(\theta)^{\sT} g(x_i,\theta) \right\}}=0.
\end{equation}

Note that, the gradient depends on the value of the Lagrange multiplier but not on the value of its gradient.



Now, Under assumption A3,  it follows that the gradient of the log empirical likelihood diverges on the set of all boundary points  $\partial\Theta_1^{(q-1)}$. more specifically the following result can be proved (see \citet[Theorem $3$]{chaudhuriMondalTeng2017}). 


\begin{thm}\label{lambda}
Let $\{ \theta^{(k)}\}$, $\theta^{(0)}\in\partial\Theta^{(0)}$ be as defined above.  Let, for each $k=1,2,\ldots$, the $\hat\lambda(\theta^{(k)})$ be the Lagrange multiplier satisfying equation \ref{eq:lagmult} with $\theta=\theta^{(k)}$.  Then
\begin{enumerate}
\item[(i)]  As $\theta^{(k)}\rightarrow \theta^{(0)}$, $\parallel\hat \lambda(\theta^{(k)})\parallel\to\infty$,
\item[(ii)] Under \text{A3} as $\theta^{(k)}\rightarrow \theta^{(0)}$, ${\parallel \nabla \log L(\theta^{(k)}) \parallel}\to \infty$.
\end{enumerate}
\end{thm}


From Theorem~\ref{lambda} it follows that at every boundary point $\theta^{(0)}$ of $\Theta_1$ such that $0$ belongs to one of the $(q-1)$-dimensional faces of $\C(x,\theta^{(0)})$, at least one component of the gradient of log empirical likelihood  that diverges to positive or negative infinity.
The gradient of the negative log empirical likelihood represents the direction of steepest increase of the negative log empirical likelihood. Since the value of the log empirical likelihood should be typically be highest around the centre of the support $\Theta_1$.  
The gradient near the boundary of $\Theta_1$ should point towards its centre.  This property can be exploited in forcing candidates of $\theta$ generated by HMC proposals to bounce back towards the interior of $\Theta_1$ from its boundaries and in consequence reducing the chance of them getting out of the support. 


\subsection{Hamiltonian Monte Carlo Sampling for Bayesian Empirical Likelihood}
\label{sec:hmc}
Hamiltonian Monte Carlo algorithm is a Metropolis algorithm where the successive steps are proposed by using a Hamiltonian dynamics. One can visualise this dynamics as a cube sliding without friction under gravity in a bowl with smooth surface.  
The total energy of the cube is the sum of the potential energy $U(\theta)$, defined by its position $\theta$ (in this case its height) and kinetic energy $K(p)$, which is determined by its momentum $p$.  
The total energy of the cube will be conserved and it will continue to slide up and down on the smooth surface of the bowl forever. The potential and the kinetic energy would however would vary with the position of the cube.

In order to use the Hamiltonian dynamics to sample from the posterior $\Pi\left(\theta\mid x\right)$ we set our potential and kinetic energy as follows:
\[
U(\theta)=-\log\Pi(\theta|x)\quad\text{and}\quad K(p)=\frac{1}{2}p^TM^{-1}p.
\]
Here, the momentum vector $p=\left(p_1,p_2,\ldots,p_d\right)$ is a totally artificial construct usually generated from a $N(0, M)$ distribution.  Most often the covariance matrix $M$ is chosen to be a diagonal matrix with diagonal $(m_1,m_2,\ldots,m_d)$, in which case each $m_i$ is interpreted as the mass of the $ith$ parameter.  The Hamiltonian of the system is the total energy
\begin{equation}\label{hamiltonian dynamics}
\mathcal{H}(\theta,p)=U(\theta)+K(p).
\end{equation}

In Hamiltonian mechanics, the variation in the position $\theta$ and momentum $p$ with time $t$ is determined by the partial derivatives of $\mathcal{H}$ with $p$ and $\theta$ respectively.  In particular,  the motion is governed by the pair of so called Hamiltonian equations:
\begin{eqnarray}
\frac{d\theta}{dt}&=&\frac{\partial \mathcal{H}}{\partial p}=M^{-1}p, \label{PDE1}\\
\frac{dp}{dt}&=&-\frac{\partial\mathcal{ H}}{\partial \theta}=-\frac{\partial U(\theta)}{\partial \theta}.\label{PDE2}
\end{eqnarray}
It is easy to show that \citep{neal2011mcmc} Hamiltonian dynamics is reversible, invariant and volume preserving, which make it suitable for MCMC sampling schemes.

In HMC we propose successive states by solving the Hamiltonian equations \eqref{PDE1} and \eqref{PDE2}.  Unfortunately, they cannot be solved analytically (except of course for a few simple cases), and they must be approximated numerically at discrete time points.  There are several ways to numerically approximate these two equations in the literature \citep{leimkuhler2004simulating}.  
For the purpose of MCMC sampling we need a method which is reversible and volume preserving. 

Leapfrog integration \citep{birdsall2004plasma} is one such method to numerically integrate the pair of Hamiltonian equations. In this method, A step-size $\epsilon$ for the time variable $t$ is first chosen. Given the value of $\theta$ and $p$ at the current time point $t$ (denoted here by $\theta(t)$ and $p(t)$ respectively), the leapfrog updates the position and the momentum at time $t+\epsilon$ as follows
\begin{eqnarray}
p(t+\frac{\epsilon}{2})&=&p(t)-\frac{\epsilon}{2}\frac{\partial U(\theta(t))}{\partial\theta},\label{leapfrog1}\\
\theta(t+\epsilon)&=&\theta(t)+\epsilon M^{-1}p(t+\frac{\epsilon}{2}),\label{leapfrog2}\\
p(t+\epsilon)&=&p(t+\frac{\epsilon}{2})-\frac{\epsilon}{2}\frac{\partial U(\theta(t+\epsilon))}{\partial\theta}.\label{leapfrog3}
\end{eqnarray}

Theoretically, due to its symmetry, the leapfrog integration satisfies the reversibility and preserves the volume. However, because of the numerical inaccuracies the volume is not preserved. This is similar to the Langevin-Hastings algorithm \citep{besag2004markov}, which is a special case of HMC.  
Fortunately, lack of invariance in volume is easily corrected. The accept-reject step in MCMC procedure ensures that the chain converges to the correct posterior.

At the beginning of each iteration of the HMC algorithm, the momentum vector $p$ is randomly sampled from the $N(0,M)$ distribution. Starting with the current state $(\theta,p)$, leapfrog integrator described above is used to simulate Hamiltonian dynamics for $T$ steps with step size of $\epsilon$. At the end of this $T$-step trajectory, the momentum $p$ is negated so that the Metropolis proposal is symmetric. At the end of this T-step iterations, the proposed state $(\theta^*,p^*)$ is accepted with probability 
\[
\min\{1,\exp(-\mathcal{H}(\theta^*,p^*)+\mathcal{H}(\theta,p))\}.
\]


The gradient of the log-posterior used in the leapfrog is a sum of the gradient of the log empirical likelihood and the gradient of the log prior.  The prior is user specified and it is hypothetically possible that even though at least one component of the gradient of the log empirical likelihood diverges at the boundary $\partial\Theta_1$, 
the log prior gradient may behave in a way so that the effect is nullified and the log posterior gradient remains finite over the closure of $\Theta_1$.  We make the following assumption on the prior mainly to avoid this possibility (see \citet{chaudhuriMondalTeng2017} for more details). 



\begin{itemize}
\item[(A4)]  Consider a sequence  $\{\theta^{(k)} \}$,  $k=1, 2,\ldots$,  of points in $\Theta_1$ such that $\theta^{(k)}$ converges to a boundary point $\theta^{(0)}$ of  $\Theta_1$. Assume that $\theta^{(0)}$  lies within $\Theta$ and $L(\theta^{(k)})$ strictly decreases to $L(\theta^{(0)})$, Then,  for some constant  $b(n, \theta^{(0)}) > -1$, we have
\begin{equation}\label{eq:liminf2}
 \liminf_{k\to\infty}   \frac{ \log\pi(\theta^{(k-1)})  -  \log\pi(\theta^{(k)}) }{ \log L(\theta^{(k-1)} ) - \log L(\theta^{(k)} ) } \ge b(n, \theta^{(0)}). 
\end{equation}
\end{itemize}

The assumption implies that near the boundary of the support, main contribution in the gradient of log-posterior with respect to any parameter appearing in the argument of the estimating equations comes from the corresponding gradient of log empirical likelihood. This is in most cases expected, especially if the sample size is large. For large sample size, the log likelihood should be the dominant term in the log-posterior. We are just assuming here that the gradients behave the same way. It would also ensure that in the boundary the gradient of the log likelihood and the log-posterior does not cancel each other, which is crucial for the proposed Hamiltonian Monte Carlo to work. 

Under these assumptions, \citet{chaudhuriMondalTeng2017} show that gradient of the log-posterior diverges to along almost every sequence as the parameter values approaches to the boundary $\partial \Theta_1$ from the interior of the support. More formally, they prove the following result. 


\begin{thm}\label{postdiv}
Consider a sequence  $\{\theta^{(k)} \}$,  $k=1, 2,\ldots$,  of points in $\Theta_1$ such that $\theta^{(k)}$ converges to a boundary point $\theta^{(0)}$ in  $\partial \Theta_1^{(q-1)}$. Furthermore, let $\theta^{(0)}$ lie within $\Theta$ and Assumption (A4) hold. Then 
\begin{equation}\label{eq:postdiv}
\Bigl\| \nabla \log \pi(\theta^{(k)} \mid x) \Bigr\| \rightarrow \infty, \hspace{.1in} \mbox{ as } \hspace{.1in} k \rightarrow \infty. 
\end{equation}
\end{thm}

Unless chosen on purpose, a sequence of points from the interior to the boundary would converge to a point on $\partial \Theta_1^{(q-1)}$ with probability $1$.  Thus under our assumptions, the gradient of the log-posterior would diverge to infinity for these sequences with a high probability. The lower dimensional faces of the convex hull (a polytope) are intersection of $q-1$ dimensional faces. It is not clear if the norm of the gradient of the posterior will diverge on those faces.  
It is conjecture that this would happen.  However, even if the conjecture is not true,  from the setup, it is clear that the sampler would rarely move to the region where the origin belongs to the lower dimensional faces of the convex hull.



\section{Package description}\label{sec:package}
The main function of the package is \code{ELHMC}. It draws samples from a empirical likelihood Bayesian posterior of the parameter of interest using Hamiltonian Monte Carlo once the estimating equations involving the parameters, the prior distribution of the parameters, the gradients of the estimating equations and the log priors are specified.  Some other parameters which controls the HMC process can also be specified.

Suppose that the data set consists of observations $ x = \left( x_1, ..., x_n \right) $ where each $ x_i $ is a vector of length $ p $ and follows a probability distribution $ F $ of family $\mathcal{F}_{\theta} $. Here $ \theta = \left(\theta_1,...,\theta_d\right) $ is the $d-$dimensional parameter of interest associated with $ F $.  Suppose there exist smooth functions $ g\left(\theta, x_i\right) = \left(g_1\left(\theta, x_i\right),...,g_q\left(\theta, x_i\right)\right)^T $ which satisfy $E_F\left[g\left(\theta,x_i\right)\right] = 0 $.  As we have explained above, \code{ELHMC} is used to draw samples of $ \theta $ from its posterior defined by an empirical likelihood.

Table \ref{arguments} enlists the full list of arguments for \code{ELHMC}. Arguments \code{data} and \code{fun} define the problem. They are the data set $ x $ and the collection of smooth functions in $ g $. The user specified starting point for $\theta$ is given in \code{initial}, whereas, \code{n.samples} is the number of samples of $\theta$ to be drawn. The gradient matrix of $g$ with respect to the parameter $\theta$ (i.e. $\nabla_{\theta}g$) has to be specified in \code{dfun}.  At the moment the function does not compute the gradient numerically by itself.  The prior \code{prior} represents the joint density functions of $ \theta_1,..,\theta_q $, which for the purpose of this description we denote by $\pi$. 
The gradient of the log prior function is specified in \code{dprior}. 
The function returns a vector containing the values of $ \frac{\partial}{\partial \theta_1}\pi\left(\theta\right),..,\frac{\partial}{\partial\theta_d}\pi\left(\theta\right) $.  Finally, the arguments \code{epsilon}, \code{lf.steps}, \code{p.variance} and \code{tol} are hyper-parameters which controls the Hamiltonian Monte Carlo algorithm.

The function \code{ELHMC} returns a list. If argument \code{detailed} is set to \code{FALSE}, the list contains samples of the parameters of interest $ \theta $, the Monte Carlo acceptance rate as listed in table \ref{returned}. If \code{detailed} is set to \code{TRUE}, additional information such as the trajectories of $ \theta $ and the momentum is included in the returned list (see Table \ref{detailedreturned}).

\begin{table}[t]
\begin{tabularx}{\textwidth}{l X}
    \hline
    \code{initial} & A vector containing the initial values of the parameter \\
    \code{data} & A matrix containing the data \\
    \code{fun} & The estimating function $ g $. It takes in a parameter vector \code{params} as the first argument and a data point vector \code{x} as the second parameter. This function returns a vector. \\
    \code{dfun} & A function that calculates the gradient of the estimating function $ g $. It takes in a parameter vector \code{params} as the first argument and a data point vector \code{x} as the second argument. This function returns a matrix. \\
    \code{prior} & A function with one argument \code{x} that returns a vector containing the prior densities of the parameters \\
    \code{dprior} & A function with one argument \code{x} that returns a vector containing the log density gradients of the parameters \\
    \code{n.samples} & Number of samples to draw \\
    \code{lf.steps} & Number of leap frog steps in each Hamiltonian Monte Carlo update \\
    \code{epsilon} & The leap frog step size \\
    \code{tol} & EL tolerance \\
    \code{detailed} & If this is set to \code{TRUE}, the function will return a list with extra information. \\
    \hline
\end{tabularx}
\caption{Arguments for function \code{ELHMC}} \label{arguments}
\end{table}

\begin{table}[b]
\begin{tabularx}{\textwidth}{l X}
    \hline
    \code{samples} & A matrix containing the parameter samples \\
    \code{acceptance.rate} & The acceptance rate \\
    \code{call} & The matched call \\
    \hline
\end{tabularx}
\caption{Elements of the list returned by \code{ELHMC} if \code{detailed = FALSE}} \label{returned}
\end{table}

\begin{table}[ht]
\begin{tabularx}{\textwidth}{l X}
    \hline
    \code{samples} & A matrix containing the parameter samples \\
    \code{acceptance.rate} & The acceptance rate \\
    \code{proposed} & A matrix containing the proposed values at \code{n.samaples - 1} Hamiltonian Monte Carlo updates \\
    \code{acceptance} & A vector of \code{TRUE/FALSE} values indicates whether each proposed value is accepted \\
    \code{trajectory} & A list with 2 elements \code{trajectory.q} and \code{trajectory.p}. These are lists of matrices containing position and momentum values along trajectory in each Hamiltonian Monte Carlo update. \\
    \code{call} & The matched call \\
    \hline
\end{tabularx}
\caption{Elements of the list returned by \code{ELHMC} if \code{detailed = TRUE}} \label{detailedreturned}
\end{table}

\section{Examples}\label{sec:examples}

In this section, we present two examples of usage of the package. Both examples in some sense supplement the conditions considered by \citet{chaudhuriMondalTeng2017}.  In each case it is seen that the function can sample from the resulting empirical likelihood based posterior quite efficiently.

\subsection{Sample the mean of a simple data set}

In the first example, suppose the data set consists of eight data points $ v = \left(v_1,...,v_8\right) $:

\begin{Schunk}
\begin{Sinput}
R> v <- rbind(c(1, 1), c(1, 0), c(1, -1), c(0, -1),
+            c(-1, -1), c(-1, 0), c(-1, 1), c(0, 1))
R> print(v)
\end{Sinput}
\begin{Soutput}
     [,1] [,2]
[1,]    1    1
[2,]    1    0
[3,]    1   -1
[4,]    0   -1
[5,]   -1   -1
[6,]   -1    0
[7,]   -1    1
[8,]    0    1
\end{Soutput}
\end{Schunk}

The parameters of interest is the mean $ \theta = \left(\theta_1, \theta_2\right) $. Since $ E\left[\theta - v_i\right] = 0$, the smooth function is $ g = \theta - v_i $ with $ \nabla_\theta g = \left(\left(1, 0\right), \left(0, 1\right)\right) $:

\begin{Schunk}
\begin{Sinput}
R> g <- function(params, x) {
+   params - x
+ }
R> dlg <- function(params, x) {
+   rbind(c(1, 0), c(0, 1))
+ }
\end{Sinput}
\end{Schunk}

Functions \code{g} and \code{dlg} are supplied to arguments \code{fun} and \code{dfun} in \code{ELHMC}. These two functions must have \code{params} as the first argument and \code{x} as the second. \code{params} represents a sample of $ \theta $ whereas \code{x} represents a data point $ v_i $ or a row in the matrix \code{v}. \code{fun} should return a vector and \code{dfun} a matrix whose $\left(i, j\right)$ entry is $ \frac{\partial g_i}{\partial{\theta_j}} $.

We assume that both $\theta_1$ and $\theta_2$ have independent standard normal distributions as priors. Next, we define the functions that calculate the prior densities and gradients of log prior densities as \code{pr} and \code{dpr} in the following ways:

\begin{Schunk}
\begin{Sinput}
R> pr <- function(x) {
+   exp(-t(x)%*%x / 2) / sqrt(2 * pi)
+ }
R> dpr <- function(x) {
+   -x
+ }
\end{Sinput}
\end{Schunk}

Functions \code{pr} and \code{dpr} are assigned to \code{prior} and \code{dprior} in \code{ELHMC}. \code{prior} and \code{dprior} must take in only one argument \code{x} and return a vector of the same length as $ \theta $.

We can now use \code{ELHMC} to draw samples of $ \theta $. Let us draw 1000 samples, with starting point $ \left(0.96, 0.97\right) $ using 10 leap frog steps with step size 0.1 for both $ \theta_1 $ and $ \theta_2 $ for each Hamiltonian Monte Carlo update:

\begin{Schunk}
\begin{Sinput}
R> library(elhmc)
R> set.seed(476)
R> thetas <- ELHMC(initial = c(0.9, 0.95), data = v, fun = g, dfun = dlg,
+                 prior = pr, dprior = dpr, n.samples = 1000,
+                 lf.steps = 15, epsilon = 0.06, detailed = TRUE)
\end{Sinput}
\end{Schunk}

\begin{figure}[t]
\centering
\subfigure[\label{theta}]{\resizebox{2in}{2in}{\includegraphics{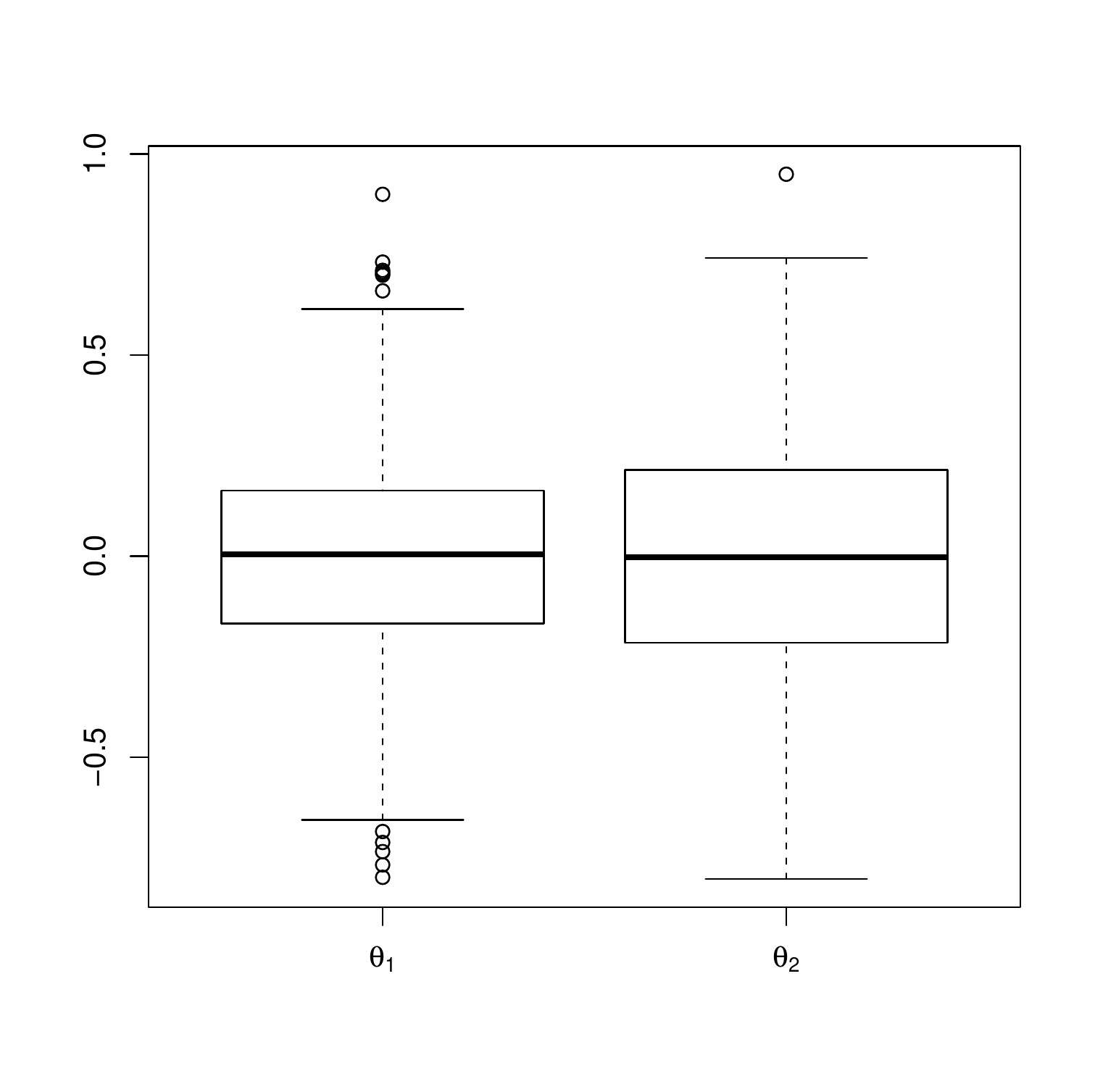}}
}\quad\subfigure[\label{trajectoryeg1}]{
\resizebox{2in}{2in}{\includegraphics{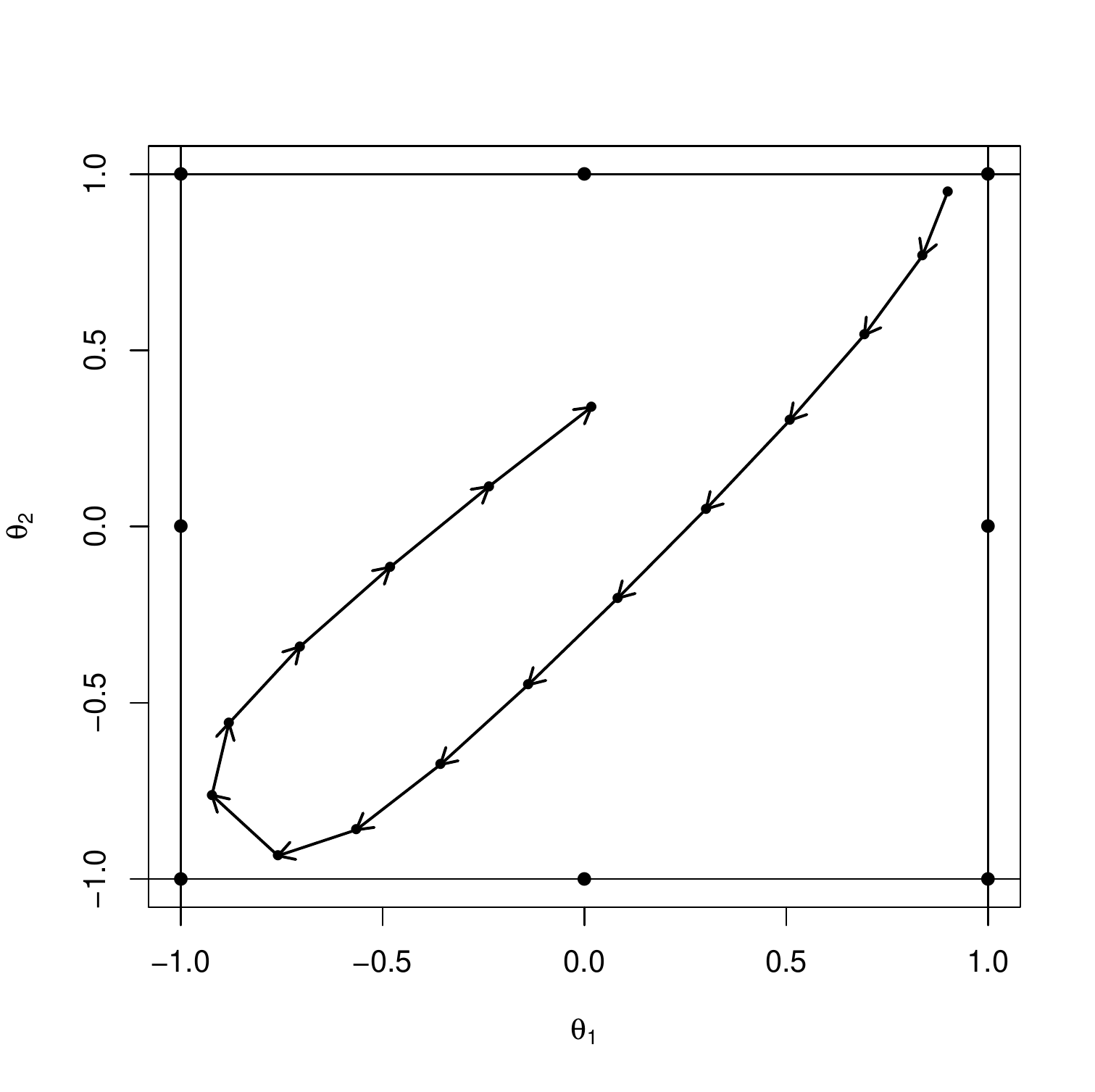}}
}
\caption{\ref{theta} Posterior distribution of $ \theta_1 $ and $ \theta_2 $ samples. \ref{trajectoryeg1} Trajectory of $ \theta $ during the first Monte Carlo update.} 
\end{figure}

We extract and visualise the distribution of the samples using a boxplot (Figure \ref{theta}):

\begin{Schunk}
\begin{Sinput}
R> boxplot(thetas$samples, names = c(expression(theta_1), expression(theta_2)))
\end{Sinput}
\end{Schunk}


Since we set \code{detailed = TRUE}, we have data on the trajectory of $ \theta $ as well as momentum $ p $. They are stored in element \code{trajectory} of \code{thetas} and can be accessed by \code{thetas$trajectory}. \code{thetas$trajectory} is a list with two elements named \code{trajectory.q} and \code{trajectory.p} denoting trajectories for $ \theta $ and momentum $ p $. \code{trajectory.q} and \code{trajectory.p} are both lists with elements \code{1}, ..., \code{n.samples - 1}. Each of these elements is a matrix containing trajectories of $ \theta $ (\code{trajectory.q}) and $ p $ (\code{trajectory.p}) at each Hamiltonian Monte Carlo update.

We illustrate by extracting the trajectories of $ \theta $ at the first update and plotting them (Figure \ref{trajectoryeg1}):

\begin{Schunk}
\begin{Sinput}
R> q <- thetas$trajectory$trajectory.q[[1]]
R> plot(q, xlab = expression(theta[1]), ylab = expression(theta[2]),
+      xlim = c(-1, 1), ylim = c(-1, 1), cex = 1, pch = 16)
R> arrows(q[-nrow(q), 1], q[-nrow(q), 2], q[-1, 1], q[-1, 2],
+        length = 0.1, lwd = 1.5)
\end{Sinput}
\end{Schunk}

The speciality in this example is in the choice of the data points in $v$.  \citet{chaudhuriMondalTeng2017} show that the chain will reflect if the one dimensional boundaries of the convex hull (in this case the unit square) has two observations, which happens with probability one for continuous distributions. In this example however, there are more than one points in two one dimensional boundaries.  However, we can see that the HMC method works very well here.  

\subsection{Logistic regression with an additional constraint}
In this example, we consider a constrained logistic regression of one binary variable on another, where the expectation of the response is known.  The frequentist estimation problem using empirical likelihood was considered by \citet{chaudhuri_handcock_rendall_2008}.  
It has been shown that empirical likelihood based formulation has major applicational advantage over the fully parametric formulation.  Below we consider a Bayesian extension of the proposed empirical likelihood based formulation and use \code{ELHMC} to sample from the resulting posterior.  

The data set $ v $ consists of $ n $ observations of two variables and two columns $ X $ and $ Y $. In the ith row $ y_i $ represents the indicator of whether a woman gave birth between time $ t - 1 $ and $ t $ while $ x_i $ is the indicator of whether she had at least one child at time $ t - 1 $.  In addition, it was known that the prevalent general fertility rate in the population was $0.06179$.
\footnote{The authors are grateful to Prof. Michael Rendall, Department of Sociology, University of Maryland, College Park, for kindly sharing the data on which this example is based.}

We are interested in fitting a logistic regression model to the data with $ X $ as the independent variable and $ Y $ the dependent variable. However, we also would like to constrain the sample general fertility rate to its value in the population.  The logistic regression model takes the form of:

$$
  P \left(Y = 1 | X = x\right) = \frac{\exp\left(\beta_0 + \beta_1 x\right)}{1 + \exp\left(\beta_0 + \beta_1 x\right)}.
$$

From the model it is clear that:

\begin{align}
E\left[y_i - \frac{\exp\left(\beta_0 + \beta_1 x_i\right)}{1 + \exp\left(\beta_0 + \beta_1 x_i\right)}\right] &= 0,\nonumber\\
E\left[x_i\left[y_i - \frac{\exp\left(\beta_0 + \beta_1 x_i\right)}{1 + \exp\left(\beta_0 + \beta_1 x_i\right)}\right]\right] &= 0.
\end{align}

Furthermore from the definition of general fertility rate we get:

$$
E\left[y_i - 0.06179\right] = 0.
$$

Following \citet{chaudhuri_handcock_rendall_2008}, we define the estimating equations $ g $ as follows:

$$
g \left(\beta, v_i\right) = \begin{bmatrix}
y_i - \frac{\exp\left(\beta_0 + \beta_1 x_i\right)}{1 + \exp\left(\beta_0 + \beta_1 x_i\right)} \\
x_i\left[y_i - \frac{\exp\left(\beta_0 + \beta_1 x_i\right)}{1 + \exp\left(\beta_0 + \beta_1 x_i\right)}\right] \\
y_i - 0.06179 \\
\end{bmatrix}
$$

The gradient of $g$ with respect to $ \beta $ is given by:

$$
\nabla_{\beta}g = \begin{bmatrix}
\frac{-\exp\left(\beta_0 + \beta_1 x_i\right)}{\left(\exp\left(\beta_0 + \beta_1 x_i\right) + 1\right)^2} & \frac{-\exp\left(\beta_0 + \beta_1 x_i\right) x_i}{\left(\exp\left(\beta_0 + \beta_1 x_i\right) + 1\right)^2} \\
\frac{-\exp\left(\beta_0 + \beta_1 x_i\right) x_i}{\left(\exp\left(\beta_0 + \beta_1 x_i\right) + 1\right)^2} & \frac{-\exp\left(\beta_0 + \beta_1 x_i\right) x_i^2}{\left(\exp\left(\beta_0 + \beta_1 x_i\right) + 1\right)^2} \\
0 & 0 \\
\end{bmatrix}
$$

In \proglang{R}, we create functions \code{g} and \code{dlg} to represent $ g $ and $ \nabla_{\beta}g $:

\begin{Schunk}
  \begin{Sinput}
R> g <- function(params, X) {
+  result <- matrix(0, nrow = nrow(X), ncol = 3)
+  a <- exp(params[1] + params[2] * X[, 1])
+  a <- a / (1 + a)
+  result[, 1] <- X[, 2] - a
+  result[, 2] <- (X[, 2] - a) * X[, 1]
+  result[, 3] <- X[, 2] - 0.06179
+  result
}
R> dg <- function(params, X) {
+  result <- array(0, c(3, 2, nrow(X)))
+  a <- exp(params[1] + params[2] * X[, 1])
+  a <- -a / (a + 1) ^ 2
+  result[1, 1, ] <- a
+  result[1, 2, ] <- result[1, 1, ] * X[, 1]
+  result[2, 1, ] <- result[1, 2, ]
+  result[2, 2, ] <- result[1, 2, ] * X[, 1]
+  result[3, , ] <- 0
+  result
}
\end{Sinput}
\end{Schunk}

We choose independent $ N\left(0, 100\right) $ priors for both $ \beta_0 $ and $ \beta_1 $:

\begin{Schunk}
\begin{Sinput}
R> pr <- function(x) {
+   exp(-0.5 * t(x)%*%x / 10 ^ 4) / sqrt(2 * pi) / 10 ^ 2
+ },
R> dpr <- function(x) {
+   -x * 10 ^ - 4
+ },
\end{Sinput}
\end{Schunk}
where \code{pr} is the prior and \code{dpr} is the gradient of the log prior for $ \beta $.

Our goal is to use \code{ELHMC} to draw samples of $ \beta = \left( \beta_0, \beta_1 \right)$ from their resulting posterior based on empirical likelihood.

We start our sampling from the $(-3.2,0.55)$ and use two stages of sampling.  In the first stage $50$ point are sampled with $\epsilon=0.001$, $T=15$ and the momentum generated from a $N(0,0.02\cdot I_2)$ distribution.  The acceptance rate at this stage is very high but it is designed to find a good starting point for the second stage, where the acceptance rate can be easily controlled.   

\begin{Schunk}
  \begin{Sinput}
R> bstart.init=c(-3.2,.55)
R> betas.init <- ELHMC(initial = bstart.init, data = data, FUN = g, DFUN = dg,
+               n.samples = 50, prior = pr, dprior = dpr, epsilon = 0.001,
+               lf.steps = 15, detailed = T, p.variance = 0.2)
  \end{Sinput}
\end{Schunk}

\begin{figure}[t]
  \centering
  \subfigure[ACF Plots. \label{fig:acf}]{
    \resizebox{2.5in}{2.5in}{\includegraphics{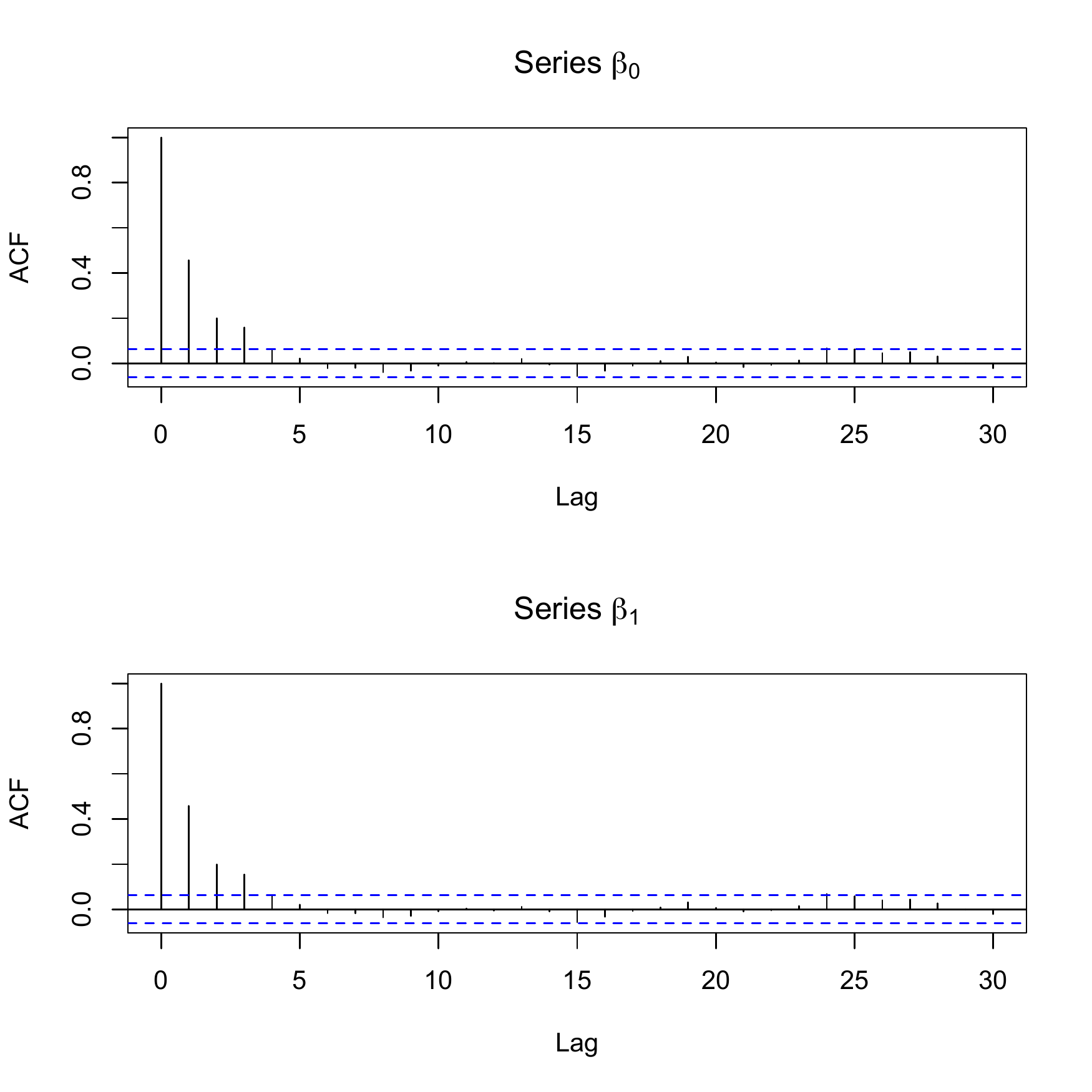}}
  }\subfigure[Density Plot. \label{fig:d}]{
    \resizebox{2.5in}{2.5in}{\includegraphics{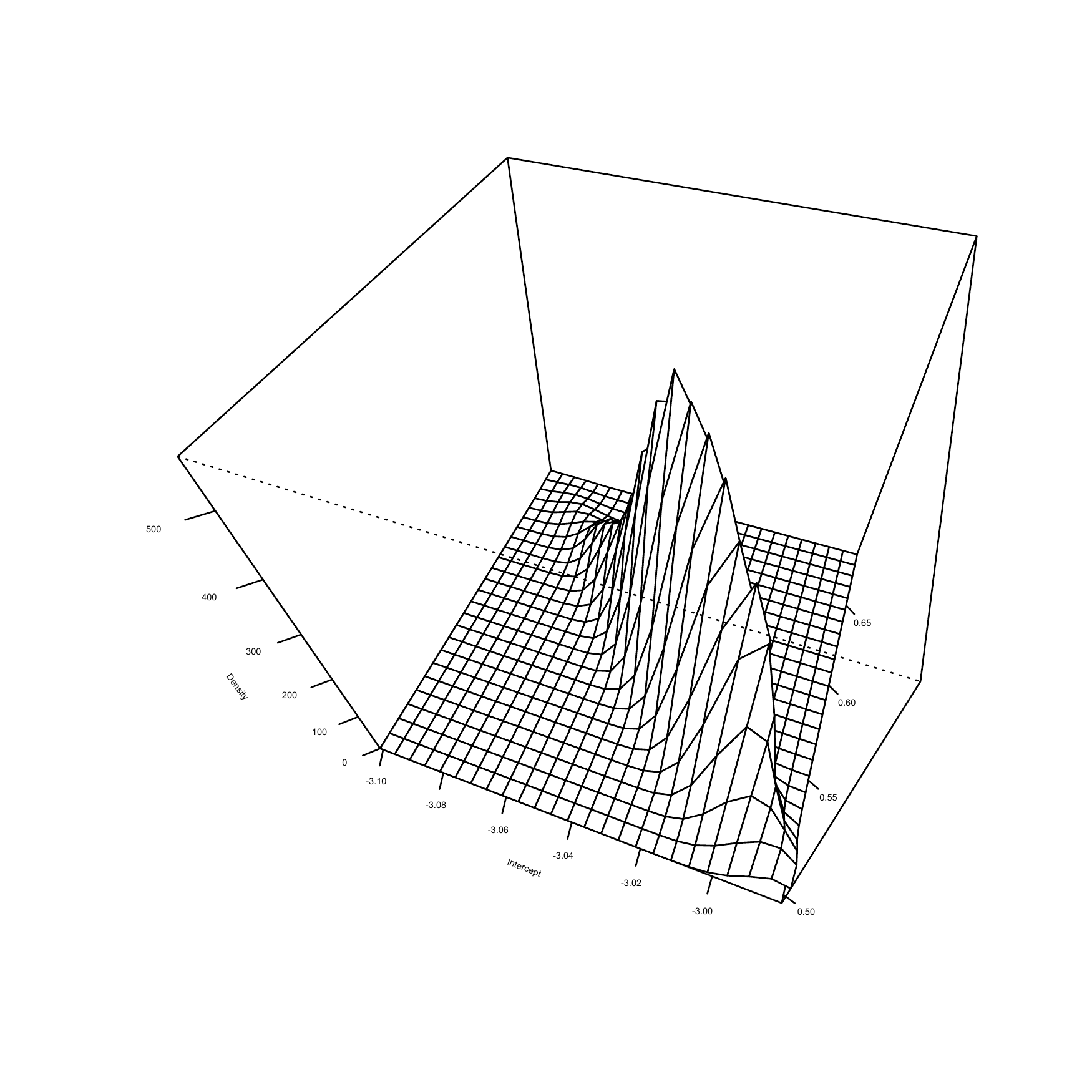}}
  }
  \caption{The autocorrelation function \ref{fig:acf} and density plot \ref{fig:d} of the samples drawn from the posterior of $ \beta $.}
  \label{fig:density}
\end{figure}

In this second stage, we draw 2000 samples of $ \beta $ with starting values as the last value from the first stage. Number of leaf frog steps per Monte Carlo update is set to 30, with step size of 0.004 for both $ \beta_0 $ and $ \beta_1 $. We use $ N \left(0, 0.02\textbf(I_2)\right) $ as prior for the momentum. 

\begin{Schunk}
\begin{Sinput}
R> bstart=betas.init$samples[50,]
R> betas <- ELHMC(initial = bstart, data = data, fun = g, dfun = dg,
+                n.samples = 2000, prior = pr, dprior = dpr, epsilon = 0.004,
+                lf.steps = 30, detailed = TRUE, p.variance = 0.02)
\end{Sinput}
\end{Schunk}

Based on our output, we can make inference about $\beta$.  As for example, the autocorrelation plots and the density plot of last $1000$ samples of $ \beta $ is shown in Figure \ref{fig:density}.

\begin{Schunk}
\begin{Sinput}
R> library(MASS)
R> beta.density <- kde2d(betas$sample[, 1], betas$samples[, 2])
R> persp(beta.density, phi = 50, theta = 20,
+       xlab = 'Intercept', ylab = '', zlab = 'Density',
+       ticktype = 'detailed', cex.axis = 0.35, cex.lab = 0.35, d = 0.7)
R> acf(betas$sample[round(n.samp/2):n.samp, 1],
+       main=expression(paste("Series ",beta[0])))
R> acf(betas$sample[round(n.samp/2):n.samp, 2],
+       main=expression(paste("Series ",beta[1])))
\end{Sinput}
\end{Schunk}

It is well known \citep{chaudhuri_handcock_rendall_2008} that the constrained estimates of $\beta_0$ and $\beta_1$ have very low standard error.  The acceptance rate is close to $78\%$.  It is evident that our software can sample from such a narrow ridge at ease, furthermore, the autocorrelation of the samples seems to decrease very quickly with the lag, which would not be the case for most other MCMC procedures.  
  
\section*{Acknowledgement}
Dang Trung Kien would like to acknowledge the support of MOE AcRF R-155-000-140-112 from National University of Singapore.  Sanjay Chaudhuri acknowledges the partial support from MOE AcRF R-155-000-192-114 from National University of Singapore.  
Authors are grateful to Professor Michael Rendall, Department of Sociology, University of Maryland, College Park for kindly sharing the data set on which the second example is based.

\begin{thebibliography}{19}
\newcommand{\enquote}[1]{``#1''}
\providecommand{\natexlab}[1]{#1}
\providecommand{\url}[1]{\texttt{#1}}
\providecommand{\urlprefix}{URL }
\expandafter\ifx\csname urlstyle\endcsname\relax
  \providecommand{\doi}[1]{doi:\discretionary{}{}{}#1}\else
  \providecommand{\doi}{doi:\discretionary{}{}{}\begingroup
  \urlstyle{rm}\Url}\fi
\providecommand{\eprint}[2][]{\url{#2}}

\bibitem[{Bergsma \emph{et~al.}(2012)Bergsma, Croon, van~der Ark
  \emph{et~al.}}]{bergsma2012empty}
Bergsma W, Croon M, van~der Ark LA, \emph{et~al.} (2012).
\newblock \enquote{The empty set and zero likelihood problems in maximum
  empirical likelihood estimation.}
\newblock \emph{Electronic Journal of Statistics}, \textbf{6}, 2356--2361.

\bibitem[{Besag(2004)}]{besag2004markov}
Besag J (2004).
\newblock \enquote{Markov chain Monte Carlo methods for statistical inference.}

\bibitem[{Birdsall and Langdon(2004)}]{birdsall2004plasma}
Birdsall CK, Langdon AB (2004).
\newblock \emph{Plasma physics via computer simulation}.
\newblock CRC Press.

\bibitem[{Boyd and Vandenberghe(2004)}]{boyd2004convex}
Boyd SP, Vandenberghe L (2004).
\newblock \emph{Convex optimization}.
\newblock Cambridge university press.

\bibitem[{Carpenter \emph{et~al.}(2017)Carpenter, Gelman
  \emph{et~al.}}]{stan2017}
Carpenter B, Gelman A, \emph{et~al.} (2017).
\newblock \enquote{Stan: A Probabilistic Programming Language.}
\newblock \emph{Journal of Statistical Software, Articles}, \textbf{76}(1),
  1--32.
\newblock ISSN 1548-7660.
\newblock \doi{10.18637/jss.v076.i01}.
\newblock \urlprefix\url{https://www.jstatsoft.org/v076/i01}.

\bibitem[{Chaudhuri \emph{et~al.}(2008)Chaudhuri, Handcock, and
  Rendall}]{chaudhuri_handcock_rendall_2008}
Chaudhuri S, Handcock MS, Rendall MS (2008).
\newblock \enquote{Generalized linear models incorporating population level
  information: an empirical-likelihood-based approach.}
\newblock \emph{Journal of the Royal Statistical Society series B},
  \textbf{70}, 311--328.

\bibitem[{Chaudhuri \emph{et~al.}(2017)Chaudhuri, Mondal, and
  Yin}]{chaudhuriMondalTeng2017}
Chaudhuri S, Mondal D, Yin T (2017).
\newblock \enquote{Hamiltonian Monte Carlo sampling in Bayesian empirical
  likelihood computation.}
\newblock \emph{Journal of the Royal Statistical Society: Series B (Statistical
  Methodology)}, \textbf{79}(1), 293--320.
\newblock ISSN 1467-9868.
\newblock \doi{10.1111/rssb.12164}.
\newblock \urlprefix\url{http://dx.doi.org/10.1111/rssb.12164}.

\bibitem[{Chen \emph{et~al.}(2008)Chen, Variyath, and
  Abraham}]{chen2008adjusted}
Chen J, Variyath A, Abraham B (2008).
\newblock \enquote{Adjusted empirical likelihood and its properties.}
\newblock \emph{Journal of Computational and Graphical Statistics},
  \textbf{17}(2), 426--443.

\bibitem[{Emerson \emph{et~al.}(2009)Emerson, Owen
  \emph{et~al.}}]{emerson2009calibration}
Emerson SC, Owen AB, \emph{et~al.} (2009).
\newblock \enquote{Calibration of the empirical likelihood method for a vector
  mean.}
\newblock \emph{Electronic Journal of Statistics}, \textbf{3}, 1161--1192.

\bibitem[{Geman and Geman(1984)}]{geman1984stochastic}
Geman S, Geman D (1984).
\newblock \enquote{Stochastic relaxation, Gibbs distributions, and the Bayesian
  restoration of images.}
\newblock \emph{Pattern Analysis and Machine Intelligence, IEEE Transactions
  on}, (6), 721--741.

\bibitem[{Grend{\'a}r and Judge(2009)}]{grendar2009empty}
Grend{\'a}r M, Judge G (2009).
\newblock \enquote{Empty set problem of maximum empirical likelihood methods.}
\newblock \emph{Electronic Journal of Statistics}, \textbf{3}, 1542--1555.

\bibitem[{Haario \emph{et~al.}(1999)Haario, Saksman, and
  Tamminen}]{haario1999adaptive}
Haario H, Saksman E, Tamminen J (1999).
\newblock \enquote{Adaptive proposal distribution for random walk Metropolis
  algorithm.}
\newblock \emph{Computational Statistics}, \textbf{14}(3), 375--396.

\bibitem[{Leimkuhler and Reich(2004)}]{leimkuhler2004simulating}
Leimkuhler B, Reich S (2004).
\newblock \emph{Simulating hamiltonian dynamics}, volume~14.
\newblock Cambridge University Press.

\bibitem[{Liu \emph{et~al.}(2010)Liu, Chen \emph{et~al.}}]{liu2010adjusted}
Liu Y, Chen J, \emph{et~al.} (2010).
\newblock \enquote{Adjusted empirical likelihood with high-order precision.}
\newblock \emph{The Annals of Statistics}, \textbf{38}(3), 1341--1362.

\bibitem[{Neal(2011)}]{neal2011mcmc}
Neal R (2011).
\newblock \enquote{MCMC for Using Hamiltonian Dynamics.}
\newblock \emph{Handbook of Markov Chain Monte Carlo}, pp. 113--162.

\bibitem[{Qin and Lawless(1994)}]{qin1994empirical}
Qin J, Lawless J (1994).
\newblock \enquote{Empirical likelihood and general estimating equations.}
\newblock \emph{The Annals of Statistics}, pp. 300--325.

\bibitem[{Tsao(2013)}]{tsao2013extending}
Tsao M (2013).
\newblock \enquote{Extending the empirical likelihood by domain expansion.}
\newblock \emph{Canadian Journal of Statistics}, \textbf{41}(2), 257--274.

\bibitem[{Tsao and Wu(2013)}]{tsao2013empirical}
Tsao M, Wu F (2013).
\newblock \enquote{Empirical likelihood on the full parameter space.}
\newblock \emph{The Annals of Statistics}, \textbf{41}(4), 2176--2196.

\bibitem[{Tsao and Wu(2014)}]{tsaoFu2014}
Tsao M, Wu F (2014).
\newblock \enquote{Extended empirical likelihood for estimating equations.}
\newblock \emph{Biometrika}, \textbf{101}(3), 703--710.

\end{thebibliography}
\setlength{\bibsep}{0pt plus 0.3ex}

\end{document}